# Quark Diagram Analysis of Bottom Meson Decays Emitting Pseudoscalar and Vector Mesons

Maninder Kaur[†], Supreet Pal Singh and R. C. Verma

*Department of Physics, Punjabi University,*
*Patiala – 147002, India.*

*e-mail:* maninderphy@gmail.com,
spsmudahar@gmail.com
*and*
rcverma@gmail.com

### *Abstract*

This paper presents the two body weak nonleptonic decays of $B$ mesons emitting pseudoscalar ($P$) and vector ($V$) mesons within the framework of the diagrammatic approach at flavor *SU(3)* symmetry level. Using the decay amplitudes, we are able to relate the branching fractions of $B \rightarrow PV$ decays induced by both $b \rightarrow c$ and $b \rightarrow u$ transitions, which are found to be well consistent with the measured data. We also make predictions for some decays, which can be tested in future experiments.

*PACS No.*:13.25.Hw, 11.30.Hv, 14.40.Nd

[†]Corresponding author: maninderphy@gmail.com

## 1. Introduction

At present, several groups at Fermi lab, Cornell, CERN, DESY, KEK and Beijing Electron Collider etc. are working to ensure wide knowledge of the heavy flavor physics. In future, a large quantity of new and accurate data on decays of the heavy flavor hadrons is expected which calls for their theoretical analysis. Being heavy, bottom hadrons have several channels for their decays, categorized as leptonic, semi-leptonic and hadronic decays [1-2]. The *b* quark is especially interesting in this respect as it has W-mediated transitions to both first generation (*u*) and second generation (*c*) quarks. Standard model provides satisfactory explanation of the leptonic and semileptonic decays but weak hadronic decays confronts serious problem as these decays experience strong interactions interferences [3-6]. So our focus is to understand the weak hadronic decays of charm and bottom hadrons emitting *s*-wave mesons.

Fortunately, the experimental progress for weak semileptonic and nonleponic decays of the bottom mesons during the last years has been really astounding, due to which a good amount of experimental data now exists [7]. This have inspired several theoretical works on the weak decays of *B*-mesons and therefore, we plan to investigate the two- body weak hadronic decays of heavy flavor hadrons in the framework of standard model. This paper is the extension to our previous work, where we had studied the $B \to PP$ decays [8]. The task is hampered for the computation of matrix elements between the initial and the final hadron states due to hadronization of final state quarks. In order to deal with these complicated matrix elements, usually the naïve and QCD factorization schemes [9-13] including probable final state interactions (FSI) [14-16] are employed to predict branching fractions of nonleptonic decays of *B*-mesons. For some channels of *B* decays, however, the factorization calculations appear to be in clear disagreement with current measurement. Thus data for the two-body decays of heavy flavor mesons have indicated the presence of large nonfactorizable contributions [17-24], especially for the color suppressed decays. These decays have also been studied using flavor symmetry, where various dynamical factors get lumped into a few reduced matrix elements, which are generally determined using some experimental results [25-28].

In the present work we have studied $B \to PV$ weak decays, investigating contributions arising from various quark level weak interaction processes. Due to the strong interaction interference on these processes, like FSI and nonfactorizable contributions, it is not possible to calculate their contributions reliably. For instance, weak annihilation and W-exchange contributions which are naively expected to be suppressed in comparison to the W-emission terms, may become significant due to possible nonfactorizable effects arising through soft-gluon exchange around the weak vertex. Since such effects are not calculable from the first principles, we employ the model independent Quark Diagram approach or Quark Diagram Scheme (hereafter referred to as QDS shown in Fig1:) [29-31]. In QDS, decay amplitudes (referred to as Quark amplitudes) can be expressed independently in terms of the topologies of possible quark flavor diagrams- like: a) the external W-emission diagram, b) the internal W-emission diagram, c) the W-exchange diagram, d) the W-annihilation, and e) the W-loop diagram, and parameterize their contributions to *B*-meson decays. The QDS has already been shown to be a useful technique for heavy flavor weak decays.

In section 2, we construct the weak Hamiltonian responsible for the $B \to PV$ decays. Choosing appropriate components of the weak quark level processes, we then obtain several straightforward relations among their decay amplitudes in Cabibbo-Kobayashi-Maskawa (CKM) enhanced as well as suppressed modes in section 3. In section 4, we proceed to derive corresponding relations among their branching fractions in the QDS using *SU(2)*-isospin, *SU(2)*-U spin and flavor *SU(3)* frameworks. In this paper, we have considered only those decays for which some experimental data exist to test the applicability of the QDS. Consequently, predictions of some of the decay branching fractions are also made, which can provide further tests of the scheme. Summary and discussion are given in the last section.

## 2. Weak Hamiltonian

In the standard model, based on the gauge groups $SU(3)_C \otimes SU(2)_L \otimes U(1)_Y$, the quarks couple to the W-boson through the weak current. The nonleptonic Hamiltonian has the usual current $\otimes$ current form

$$H_w = \frac{G_F}{\sqrt{2}} J^+_\mu J^\mu + h.c. \tag{1}$$

where the weak current $J_\mu$ is given by

$$J_\mu = (\bar{u}\ \bar{c}\ \bar{t})\ \gamma_\mu(1\text{-}\gamma_5) \begin{pmatrix} d' \\ s' \\ b' \end{pmatrix} . \tag{2}$$

Weak eigenstates ($d'$, $s'$ and $b'$) are related to the mass eigenstates ($d$, $s$ and $b$) through the Cabibbo-Kobayashi-Maskawa mixing. We consider hadronic decays of *B*-mesons induced at the quark level by $b \to c/u$ transitions. The weak Hamiltonian generating the *b*-quark decays is thus given by

$$\begin{aligned} H_w^{\Delta b=1} = \frac{G_F}{\sqrt{2}} [\, & V_{ub} V^*_{cd}\ (\bar{u}\, b)(\bar{d}\ c) + V_{ub} V^*_{cs}\ (\bar{u}\, b)(\bar{s}c) + V_{ub} V^*_{ud}\ (\bar{u}b)(\bar{d}u) \\ & + V_{ub} V^*_{us}\ (\bar{u}b)(\bar{s}u) + V_{cb} V^*_{ud}\ (\bar{c}b)(\bar{d}u) + V_{cb} V^*_{us}\ (\bar{c}b)(\bar{s}u) \\ & + V_{cb} V^*_{cs}\ (\bar{c}b)(\bar{s}c) + V_{cb} V^*_{cd}\ (\bar{c}b)(\bar{d}c)\, ]. \end{aligned} \tag{3}$$

The color and space-time structure is omitted. Selection rules for various decay modes generated by the Hamiltonian are given below.

| | |
|---|---|
| (i) CKM enhanced modes: | $\Delta C = 1, \Delta S = 0; \Delta C = 0, \Delta S = -1;$ |
| (ii) CKM Suppressed modes: | $\Delta C = 1, \Delta S = -1; \Delta C = 0, \Delta S = 0;$ |
| (iii) CKM doubly suppressed modes: | $\Delta C = \Delta S = -1; \Delta C = -1, \Delta S = 0.$ |

Since only quark fields appear in the Hamiltonian, the *B*-meson decays are seriously affected by the strong interactions. One usually identifies the two scales in these decays: short distance scale at which W-exchange takes place and long distance scale where final state hadrons are formed. The short distance effects are calculable using the perturbative QCD, which are expressed in terms of certain QCD coefficients. The long-distance effects being non-perturbative are the source of major problems in obtaining the decay amplitudes from the Hamiltonian, even after including the short distance modifications [32-33].

There are many ways that the quarks produced in a weak nonleptonic process can arrange themselves into final state hadrons. All *B*-meson decays can be expressed in terms of a few quark level diagrams [29-31] (shown below):

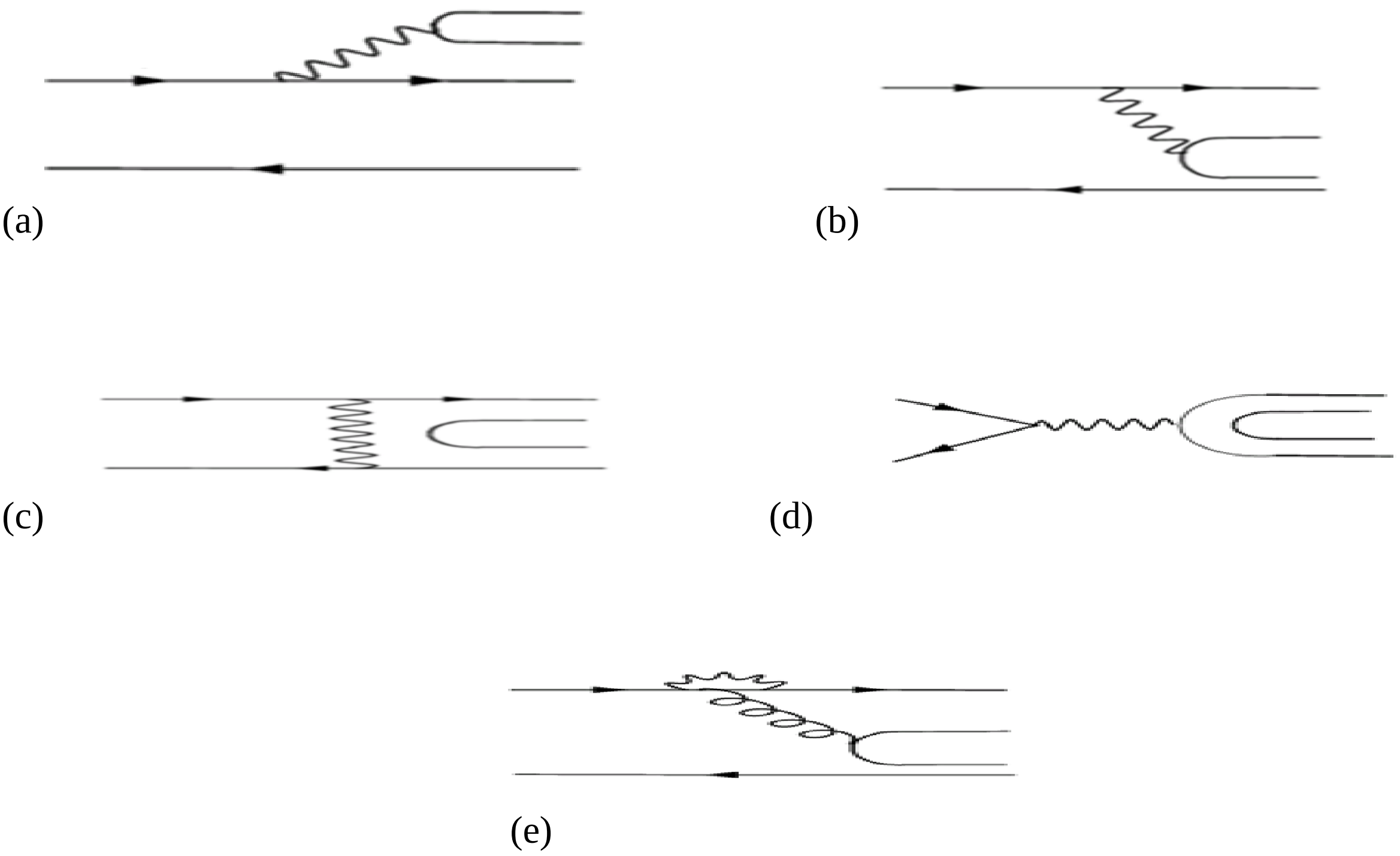

Fig1: a) the external W-emission diagram, b) the internal W-emission diagram, c) the W-exchange diagram, d) the W-annihilation, and e) the W-loop penguin diagram [29-31]

Initially, it was expected that W-exchange and W-annihilation diagrams are suppressed due to the helicity and color considerations while the penguin diagrams, involving W-loop, contribute to only two out of the six decay modes. Thus the dominant quark level process apparently seems to involve W-emission in which light quark in the *B*-meson behaves like spectator. However, measurements of some of *B*-meson decays have challenged this simple picture and it is now established that the non-spectator contributions may play a significant role in understanding the weak decays of heavy flavor hadrons. In fact, exchange of the soft gluons around the weak vertex also enhances such non-spectator contributions from the W-exchange, W-annihilation and W-loop diagrams. Unfortunately, these effects, being non-perturbative, cannot be determined unambiguously from the first principles. QCD sum rules approach has been used to estimate them, but so far it has not given reliable results.

In the absence of the exact dynamical calculations, we have employed the QDS to investigate contributions from different weak quark level diagrams (shown above). Such a scheme gives model independent way to analyze data to test the mechanism of the various quark level processes and to make useful predictions for the meson decays. The decay amplitudes are obtained using the valence quark structure of the particles involved in the *B*-meson decays. Using the tensorial notation, the decay amplitudes are then obtained from the following contractions:

$$H_w^{\Delta b=1} = [a(B^m P_m^i V_n^k) + d(B^i P_n^m V_m^k)]\, H_{[i,k]}^n + [a'(B^m P_m^i V_n^k) + d'(B^i P_n^m V_m^k)]\, H_{(i,k)}^n$$
$$+ [a_v(B^m P_n^k V_m^i) + d_v(B^i P_m^k V_n^m)]\, H_{[i,k]}^n + [a_v'(B^m P_n^k V_m^i) + d_v'(B^i P_m^k V_n^m)]\, H_{(i,k)}^n$$
$$+ [c(B^n P_n^m V_m^i)]\, H_i + [c_v(B^n P_m^i V_n^m)]\, H_i. \quad (4)$$

The brackets [,] and (,) respectively, denote antisymmetrization and symmetrization among the indices *i, k.* In the flavor *SU(4)*, the *b*-quark behaves like singlet, and *u*, *d*, *s* and *c* quarks form a quartet. Thus Hamiltonian for $\Delta b$ = 1 weak process belong to the representations appearing in

$$4^* \otimes 1 \otimes 4^* \otimes 4 = 4^* \oplus 4^* \oplus 20' \oplus 36^*. \quad (5)$$

The weak spurion $H_i$, $H_{[i,k]}^n$ and $H_{(i,k)}^n$ belong to the 4*, 20', and 36* representations respectively. However, we do not use the complete *SU(4)*-QDS due to *SU(4)* being badly broken and exploit the QDS at the *SU(3)* level through the following *SU(4)→ SU(3)* decomposition:

$$\begin{aligned} 4^* &\supset 3^* \oplus 1, \\ 20' &\supset 8 \oplus (6 \oplus 3^*) \oplus 3, \\ 36^* &\supset 6^* \oplus (15 \oplus 3^*) \oplus (8 \oplus 1) \oplus 3. \end{aligned} \quad (6)$$

The *SU(3)*-QDS relates ΔC = 1, ΔS = 0 mode with ΔC = 1, ΔS = -1**;** ΔC = 0, ΔS = -1 mode with ΔC = 0, ΔS = 0; and ΔC = ΔS = -1 mode with ΔC = -1, ΔS = 0. The tensor $B^i$ denotes the parent *B*-mesons:

$$B^1 = B^+(\bar{b}u),\ B^2 = B^0(\bar{b}d),\ B^3 = B_s^0(\bar{b}s),\ B^4 = B_c^+(\bar{b}c). \quad (7)$$

The $P_j^i$ and $V_j^i$ denote $4^* \otimes 4$ matrices of bottomless pseudoscalar and vector mesons.

$$P_j^i = \begin{pmatrix} P_1^1 & \pi^+ & K^+ & \overline{D^0} \\ \pi^- & P_2^2 & K^0 & D^- \\ K^- & \overline{K^0} & P_3^3 & D_s^- \\ D^0 & D^+ & D_s^+ & P_4^4 \end{pmatrix}, \quad V_j^i = \begin{pmatrix} V_1^1 & \rho^+ & K^{*+} & \overline{D^{*0}} \\ \rho^- & V_2^2 & K^{*0} & D^{*-} \\ K^{*-} & \overline{K^{*0}} & V_3^3 & D_s^{*-} \\ D^{*0} & D^{*+} & D_s^{*+} & V_4^4 \end{pmatrix}. \quad (8)$$

Using *SU(3)* nonet (or *SU(4)* sixteenplet) symmetry, the diagonal states are taken to be:

$$P_1^1 = \frac{\pi^0 + \eta\sin\theta_p + \eta'\cos\theta_p}{\sqrt{2}},\ P_2^2 = \frac{-\pi^0 + \eta\sin\theta_p + \eta'\cos\theta_p}{\sqrt{2}},\ P_3^3 = -\eta\cos\theta_p + \eta'\sin\theta_p,\ P_4^4 = \eta_c, \quad (9)$$

$$V_1^1 = \frac{\rho^0 + \omega}{\sqrt{2}},\ V_2^2 = \frac{-\rho^0 + \omega}{\sqrt{2}},\ V_3^3 = -\phi,\ V_4^4 = \psi. \quad (10)$$

Where $\theta_p$ governs the $\eta - \eta'$ mixing angle and is given as $\theta_p = \theta_{ideal} - \phi_P$; $\phi_P = -10^\circ$ (in case of pseudoscaler meson) follows from the radiative decay widths and $\eta_c(\bar{c}c)$ is taken to be charmonium iso-singlet (ideal mixing). On the other hand, the mixing angle for $\omega-\phi$ (vector mesons) is $\theta_V = \theta_{ideal} - \phi_V$, for ideal $\omega-\phi$ mixing $\boldsymbol{\theta_V = 0}$. $\phi(s\bar{s})$ and $\psi(c\bar{c})$ are isosinglet physical mesons [4, 7].

There exists a straight correspondence between the terms appearing in (4) and various quark level diagrams. The terms with coefficients *(a + a')* and $(a_v + a_v')$ represent external W-emission, *(a - a')* and $(a_v - a_v')$ represent internal W-emission, the terms with coefficients *(d - d')* and $(d_v - d_v')$ represent W-exchange and *(d + d')* and $(d_v + d_v')$ represents W-annihilation processes. The last term having coefficient *c* and $c_v$ represents the W-loop penguin diagram contributions. All the QCD effects have been absorbed in these parameters. In addition, the following contractions may also contribute.

$$\begin{aligned} &[h(B^i P_n^k V_m^m)]\, H_{[i,k]}^n + [h'(B^i P_n^k V_m^m)]\, H_{(i,k)}^n \\ &+[h_v (B^i P_m^m V_n^k)]\, H_{[i,k]}^n + [h_v'(B^i P_m^m V_n^k)]\, H_{(i,k)}^n \\ &+[f(B^i P_n^m V_m^n) + f'(B^i P_m^m V_n^n) + f''(B^n P_n^i V_m^m)]\, H_i \\ &+[f_v(B^i P_m^n V_n^m) + f_v'(B^i P_n^n V_m^m) + f_v''(B^n P_m^m V_n^i)]\, H_i . \end{aligned} \tag{11}$$

However, these terms correspond to OZI violating diagrams which are expected to be suppressed, and hence are ignored in the present scheme.

## 3. Decay Amplitudes Relations

Choosing the relevant components of the Hamiltonian given in (3), we obtain the decay amplitudes in the *SU(3)* QDS for $B \to PV$ for which some experimental results are available. Depending upon the weak quark level processes involved in these decays, we have categorized their relations given below in three different ways:

1) Only single weak process (W-emission or W-exchange or W-annihilation) given as in (*A*, *B, C* and *D*);
2) Combination of two weak processes (other than penguin diagram) given as in (*E*, *F*, *G* and *H*);
3) Combination of penguin diagram with other weak processes, (I, J and K).

### *A. __W-external emission__:*

$$A(B_s^0 \to \pi^+ D_s^{*-}) = V_{ud}/V_{us}\, A(B^0 \to K^+ D^{*-}) \tag{12}$$

$$A(B^0 \to D^- K^{*+}) = V_{us}/V_{ud}\, A(B_s^0 \to D_s^- \rho^+) \tag{13}$$

$$A(B^0 \to \pi^- D_s^{*+}) = \sqrt{2}\, A(B^+ \to \pi^0 D_s^{*+}) \tag{14}$$

$$A(B^+ \to D_s^+ \rho^0) = A(B^+ \to D_s^+ \omega) \tag{15}$$

$$A(B^+ \to D_s^+ \rho^0) = (1/\sqrt{2})\, A(B^0 \to D_s^+ \rho^-) \tag{16}$$

$$A(B_s^0 \to K^- D^{*+}) = V_{dc}/V_{sc}\, A(B^0 \to \pi^- D_s^{*+}) \tag{17}$$

$$A(B_s^0 \to D^+ K^{*-}) = V_{dc}/V_{sc}\, A(B^0 \to D_s^+ \rho^-) \tag{18}$$

**B. <u>W-internal emission</u>:**

$$A(B^+ \to K^+ \psi) = A(B^0 \to K^0 \psi) \qquad (19)$$

$$A(B^0 \to \eta_c K^{*0}) = A(B^+ \to \eta_c K^{*+}) \qquad (20)$$

$$A(B_s^0 \to \bar{D}^0 \bar{K}^{*0}) = (- V_{ud}/V_{us})\, A(B_s^0 \to \bar{D}^0 \varphi) \qquad (21)$$

$$A(B_s^0 \to \eta \psi) = (-\cot\theta_p)\, A(B_s^0 \to \eta' \psi) \qquad (22)$$

$$A(B_s^{0.} \to \bar{D}^0 \varphi) = - A(B^0 \to \bar{D}^0 K^{*0}) \qquad (23)$$

$$A(B^+ \to \pi^+ \psi) = (-\surd 2)\, A(B^0 \to \pi^0 \psi) \qquad (24)$$

$$A(B_s^0 \to \bar{K}^0 \psi) = V_{cd}/V_{cs}\, A(B^+ \to K^+ \psi) \qquad (25)$$

$$A(B^0 \to \eta \psi) = \tan\theta_p\, A(B^0 \to \eta' \psi) \qquad (26)$$

$$A(B_s^0 \to \eta_c \bar{K}^{*0}) = V_{cd}/V_{cs}\; A(B^+ \to \eta_c K^{*+}) \qquad (27)$$

$$A(B^+ \to \eta_c \rho^+) = V_{cd}/V_{cs}\, A(B^+ \to \eta_c K^{*+}) \qquad (28)$$

$$A(B_s^0 \to \bar{K}^0 \bar{D}^{*0}) = V_{ud}/V_{us}\, A(B^0 \to K^0 \bar{D}^{*0}) \qquad (29)$$

$$A(B^0 \to \eta_c \rho^0) = (-1/\surd 2)(V_{cd}/V_{cs})\, A(B^0 \to \eta_c K^{*0}) \qquad (30)$$

$$A(B^0 \to \eta_c \omega) = (-1/\surd 2)(V_{cd}/V_{cs})\, A(B_s^0 \to \eta_c \varphi) \qquad (31)$$

$$= (1/\surd 2)(V_{cd}/V_{cs})\, A(B^0 \to \eta_c K^{*0})$$

$$A(B_s^0 \to D^0 \bar{K}^{*0}) = V_{cd}/V_{cs}\, A(B^0 \to D^0 K^{*0}) \qquad (32)$$

**C. <u>W-annihilation only</u>:**

$$A(B^+ \to D_s^+ \bar{K}^{*0}) = (- V_{cd}/V_{cs})\; A(B^+ \to D_s^+ \varphi) \qquad (33)$$

$$A(B^+ \to \bar{K}^0 D_s^{*+}) = V_{cd}/V_{cs}\, A(B^+ \to K^0 D^{*+}) \qquad (34)$$

$$A(B^+ \to D_s^+ \bar{K}^{*0}) = V_{cd}/V_{cs}\, A(B^+ \to D^+ K^{*0}) \qquad (35)$$

**D. <u>W-exchange only</u>:**

$$A(B_s^0 \to \pi^+ D^{*-}) = V_{us}/V_{ud}\; A(B^0 \to K^+ D_s^{*-}) \qquad (36)$$

$$A(B_s^0 \to \pi^0 \bar{D}^{*0}) = (1/\surd 2)(V_{us}/V_{ud})\, A(B^0 \to K^+ D_s^{*-}) \qquad (37)$$

$$A(B_s^0 \to D^- \rho^+) = V_{us}/V_{ud}\, A(B^0 \to D_s^- K^{*+}) \qquad (38)$$

$$A(B_s^0 \to \bar{D}^0 \omega) = A(B_s^0 \to \bar{D}^0 \rho^0) = (1/\surd 2)(V_{us}/V_{ud})\, A(B^0 \to D_s^- K^{*+}) \qquad (39)$$

$$A(B_s^0 \to D^+ D^{*-}) = V_{cs}/V_{cd}\, A(B^0 \to D_s^+ D_s^{*-}) \qquad (40)$$

$$A(B_s^0 \to \bar{D}^0 \psi) = V_{ud}/V_{us}\, A(B^0 \to \bar{D}^0 \psi) \qquad (41)$$

$$A(B^0 \to K^- D_s^{*+}) = V_{dc}/V_{cs}\, A(B_s^0 \to \pi^- D^{*+}) \qquad (42)$$

$$A(B_s^0 \to \pi^0 D^{*0}) = (1/\surd 2)\, A(B_s^0 \to \pi^- D^{*+}) \qquad (43)$$

**E. W-internal emission and W-external emission:**

$$A(B^+\to \bar{D}^0K^{*+}) = V_{us}/V_{du}\, A(B^+\to \bar{D}^0 \rho^+) \quad (44)$$

$$A(B^+\to K^+ \bar{D}^{*0}) = V_{us}/V_{du}\, A(B^+\to \pi^+ \bar{D}^{*0}) \quad (45)$$

**F. W-internal emission and W-exchange:**

$$A(B^0\to\eta\, \bar{D}^{*0}) = \tan\theta_p\, A(B^0\to\eta'\, \bar{D}^{*0}) \quad (46)$$

**G. W-internal emission and W- annihilation:**

$$A(B_c^+\to \bar{D}^0 D_s^{*+}) = V_{us}/V_{du}\, A(B_c^+\to \bar{D}^0 D^{*+}) \quad (47)$$

**H. W-external emission and W-exchange:**

$$A\, (B_s^0\to D_s^- K^{*+}) = V_{us}/V_{du}\, A(B^0\to D^- \rho^+) \quad (48)$$

$$A(B_s^0\to K^+ D_s^{*-}) = V_{us}/V_{du}\, A(B^0\to \pi^+ D^{*-}) \quad (49)$$

**I. W-annihilation and Penguin:**

$$A(B^+\to\pi^+ K^{*0}) = -A(B^+\to K^+ \varphi) \quad (50)$$

$$A(B_c^+\to D^+ K^{*0}) = -A(B_c^+\to D_s^+ \varphi) \quad (51)$$

**J. W-internal, W-external &W- annihilation and Penguin:**

$$A(B^+\to K^+ \rho^0) = A(B^+\to K^+ \omega) \quad (52)$$

$$A(B^+\to\eta\, \rho^+) = \tan\theta_p\, A(B^+\to\eta'\, \rho^+) \quad (53)$$

**K. W- internal emission, W- exchange & Penguin:**

$$A(B^0\to\eta\, \rho^0) = \tan\theta_p\, A(B^0\to\eta'\, \rho^0) \quad (54)$$

Note that the relations (14, 16, 19-20, 24, 43) follow from the *SU(2)*-isospin framework, and (12-13, 17-18, 28-29, 32, 34-36, 38, 40-42, 44-45, 47-49) from the *SU(2)*- U spin for QDS.

## 4. Relations and Predictions for Branching Fractions

The decay rate formula for $B \to PV$ has the generic form:

$$\Gamma(B \to PV) = (non\text{-}kinematic\ factors)^2 \times \left( \frac{k^3}{8\pi\, m_V^2} \right) |A|^2, \quad (55)$$

where $k$ is the 3-momentum of the final states and is given by

$$k = |p_1| = |p_2| = \frac{1}{2m_B}\left\{\left(m_B^2 - \left(m_p + m_V\right)^2\right)\left(m_B^2 - \left(m_P - m_V\right)^2\right)\right\}^{1/2}. \qquad (56)$$

Several relations are obtained among branching fractions of the decays of $B^+$, $B^0$, $B_s^0$ and $B_c^+$ mesons, corresponding to the decay amplitude relations given in the previous section. We have used the available experimental values to check the consistency of the relations obtained and to predict the branching fractions of some of the decays not observed so far. We give our values just below the branching relations. These values are obtained by multiplying the known experimental value with the factor (given on RHS). For instance, in relation (57), value given for $B(B_s^0 \to \pi^+ D_s^{*-})$ is obtained by multiplying the experimental value of branching fraction $B(B^0 \to K^+ D^{*-}) = (2.14 \pm 0.16) \times 10^{-4}$ with the factor 17.59. Similar to the decay amplitude relations, we have categorized relations among the branching fractions according to the contributions arising from one or more of the weak quark diagrams. We have also distinguished the $b \to s$ penguin process from that of $b \to d$.

## A. *W-external emission:*

$B(B_s^0 \to \pi^+ D_s^{*-}) = 17.59\, B(B^0 \to K^+ D^{*-})$ (57)
$(0.20 \pm 0.05) \times 10^{-2}$ $(0.37 \pm 0.03) \times 10^{-2}$

$B(B^0 \to D^- K^{*+}) = 0.04\, B(B_s^0 \to D_s^- \rho^+)$ (58)
$(4.5 \pm 0.7) \times 10^{-4}$ $(2.7 \pm 0.6) \times 10^{-4}$

$B(B^+ \to \pi^0 D_s^{*+}) = 0.53\, B(B^0 \to \pi^- D_s^{*+})$ (59)
$< 26 \times 10^{-5}$ $(1.1 \pm 0.2) \times 10^{-5}$

$B(B^+ \to D_s^+ \rho^0) = 1.03\, B(B^+ \to D_s^+ \omega)$ (60)
$< 3 \times 10^{-4}$ $< 4.12 \times 10^{-4}$

$B(B^+ \to D_s^+ \rho^0) = 0.54\, B(B^0 \to D_s^+ \rho^-)$ (61)
$< 3 \times 10^{-4}$ $< 0.13 \times 10^{-4}$

$B(B_s^0 \to K^- D^{*+}) = 0.06\, B(B^0 \to \pi^- D_s^{*+})$ (62)
$(1.3 \pm 0.2) \times 10^{-6}$

$B(B_s^0 \to D^+ K^{*-}) = 0.04\, B(B^0 \to D_s^+ \rho^-)$ (63)
$< 0.1 \times 10^{-5}$

## B. *W-internal emission:*

$B(B^+ \to K^+ \psi) = 1.07\, B(B^0 \to K^0 \psi)$ (64)
$(1.03 \pm 0.03) \times 10^{-3}$ $(0.93 \pm 0.03) \times 10^{-3}$

$B(B^0 \to \eta_c K^{*0}) = 0.93\, B(B^+ \to \eta_c K^{*+})$ (65)
$(0.6 \pm 0.1) \times 10^{-3}$ $(0.93^{+0.47}_{-0.37}) \times 10^{-3}$

$$B(B_s^0 \to \bar{D}^0\ \bar{K}^{*0}) = 25.94\, B(B_s^0 \to \bar{D}^0\ \varphi) \qquad (66)$$
$(4.4 \pm 0.6)\times 10^{-4}$ $\qquad (7.78 \pm 2.08)\times 10^{-4}$

$$B(B_s^0 \to \eta\ \psi) = 1.09\, B(B_s^0 \to \eta'\ \psi) \qquad (67)$$
$(3.9 \pm 0.7)\times 10^{-4}$ $\qquad (3.6 \pm 0.4)\times 10^{-4}$

$$B(B_s^{0.} \to \bar{D}^0\ \varphi) = 0.76\, B(B^0 \to \bar{D}^0 K^{*0}) \qquad (68)$$
$(2.4 \pm 0.7)\times 10^{-5}$ $\qquad (3.2 \pm 0.5)\times 10^{-5}$

$$B(B^+ \to \pi^+ \psi) = 2.14\, B(B^0 \to \pi^0\ \psi) \qquad (69)$$
$(4.1 \pm 0.4)\times 10^{-5}$ $\qquad (3.76 \pm 0.34)\times 10^{-5}$

$$B(B_s^0 \to \bar{K}^0 \psi) = 0.05\, B(B^+ \to K^+ \psi) \qquad (70)$$
$(1.87 \pm 0.17)\times 10^{-5}$ $\qquad (5.13 \pm 0.16)\times 10^{-5}$

$$B(B^0 \to \eta\ \psi) = 1.44\, B(B^0 \to \eta'\ \psi) \qquad (71)$$
$(1.07 \pm 0.24)\times 10^{-5}$ $\qquad (1.1 \pm 0.3)\times 10^{-5}$

$$B(B_s^0 \to \eta_c\ \bar{K}^{*0}) = 0.05\, B(B^+ \to \eta_c K^{*+}) \qquad (72)$$
$\qquad (0.05^{+0.03}_{-0.02}) \times 10^{-3}$

$$B(B^+ \to \eta_c\, \rho^+) = 0.08\, B(B^+ \to \eta_c K^{*+}) \qquad (73)$$
$\qquad (0.08^{+0.04}_{-0.03}) \times 10^{-3}$

$$B(B_s^0 \to \bar{K}^0\ \bar{D}^{*0}) = 19.72\, B(B^0 \to K^0\ \bar{D}^{*0}) \qquad (74)$$
$\qquad (7.09 \pm 2.36)\times 10^{-4}$

$$B(B^0 \to \eta_c\, \rho^0) = 0.04\, B(B^0 \to \eta_c K^{*0}) \qquad (75)$$
$\qquad (2.33 \pm 0.33)\times 10^{-5}$

$$B(B^0 \to \eta_c\, \omega) = 0.05\, B(B_s^0 \to \eta_c\, \varphi) = 0.04\, B(B^0 \to \eta_c K^{*0}) \qquad (76)$$
$\qquad (2.33 \pm 0.33)\times 10^{-5}$

$$B(B_s^0 \to D^0\ \bar{K}^{*0}) = 0.05\, B(B^0 \to D^0 K^{*0}) \qquad (77)$$
$\qquad <0.06 \times 10^{-5}$

## C. <u>W-annihilation only:</u>

$$B(B^+ \to D_s^+\ \ \bar{K}^{*0}) = 0.07\, B(B^+ \to D_s^+\ \varphi) \qquad (78)$$
$<4.4\times 10^{-6}$ $\qquad (0.12 \pm 0.06)\times 10^{-6}$

$$B(B^+ \to \bar{K}^0 D_s^{*+}) = 0.05\, B(B^+ \to K^0 D^{*+}) \qquad (79)$$
$<90\times 10^{-5}$ $\qquad <0.04\times 10^{-5}$

$$B(B^+ \to D_s^+\ \ \bar{K}^{*0}) = 0.05\, B(B^+ \to D^+ K^{*0}) \qquad (80)$$
$<4.4\times 10^{-6}$ $\qquad <0.09\times 10^{-6}$

***D. W-exchange only:***

$$B(B_s^{\,0}\rightarrow \pi^+ D^{*-}) = 0.06\, B(B^0\rightarrow K^+ D_s^{*-}) \quad (81)$$
$$<6.1\times 10^{-6} \qquad (1.31 \pm 0.18)\times 10^{-6}$$

$$B(B_s^{\,0}\rightarrow \pi^0\ \bar{D}^{*0}) = 0.03\, B(B^0\rightarrow K^+ D_s^{*-}) \quad (82)$$
$$(0.11 \pm 0.03)\times 10^{-5}$$

$$B(B_s^{\,0}\rightarrow D^- \rho^+) = 0.08\, B(B^0\rightarrow D_s^- K^{*+}) \quad (83)$$
$$(0.3 \pm 0.1)\times 10^{-5}$$

$$B(B_s^0\rightarrow \bar{D}^0\, \omega) = B(B_s^0\rightarrow \bar{D}^0 \rho^0) = 0.04\, B(B^0\rightarrow D_s^- K^{*+}) \quad (84)$$
$$(1.4 \pm 0.4)\times 10^{-6}$$

$$B\,(B_s^0\rightarrow D^+ D^{*-}) = 27.59\, B(B^0\rightarrow D_s^+ D_s^{*-}) \quad (85)$$
$$< 3.6\times 10^{-3}$$

$$B(B_s^{\,0}\rightarrow \bar{D}^0\, \psi) = 0.07\, B(B^0\rightarrow \bar{D}^0\, \psi) \quad (86)$$
$$<0.09\times 10^{-5}$$

$$B(B^0\rightarrow K^- D_s^{*+}) = 0.042\, B(B_s^0\rightarrow \pi^- D^{*+}) \quad (87)$$
$$<0.26\times 10^{-6}$$

$$B(B_s^{\,0}\rightarrow \pi^0 D^{*0}) = 0.5\, B(B_s^0\rightarrow \pi^- D^{*+}) \quad (88)$$
$$<3.1\times 10^{-6}$$

***E. W-internal emission and W-external emission:***

$$B(B^+\rightarrow \bar{D}^0 K^{*+}) = 0.037\, B(B^+\rightarrow \bar{D}^0 \rho^+) \quad (89)$$
$$(5.3 \pm 0.4)\times 10^{-4} \qquad (4.90 \pm 0.66)\times 10^{-4}$$

$$B(B^+\rightarrow K^+\ \bar{D}^{*0}) = 0.05\, B(B^+\rightarrow \pi^+\ \bar{D}^{*0}) \quad (90)$$
$$(4.20\pm 0.34)\times 10^{-4} \qquad (2.60\pm 0.13)\times 10^{-4}$$

***F. W-internal emission and W-exchange:***

$$B(B^0\rightarrow \eta\ \bar{D}^{*0}) = 1.27\, B(B^0\rightarrow \eta'\ \bar{D}^{*0}) \quad (91)$$
$$(2.3 \pm 0.6)\times 10^{-4} \qquad (1.78 \pm 0.28)\times 10^{-4}$$

***G. W-internal emission and W- annihilation:***

$$B(B_c^+\rightarrow \bar{D}^0 D_s^{*+}) = 0.04\, B(B_c^+\rightarrow \bar{D}^0 D^{*+}) \quad (92)$$
$$<0.27\times 10^{-3}$$

***H. W-external emission and W-exchange:***

$$B\,(B_s^0\rightarrow D_s^- K^{*+}) = 0.04\, B(B^0\rightarrow D^- \rho^+) \quad (93)$$
$$(2.85\pm 0.47)\times 10^{-4}$$

$$B(B_s^0 \to K^+ D_s^{*-}) = 0.04\, B(B^0 \to \pi^+ D^{*-}) \qquad (94)$$
$$(1.21 \pm 0.05) \times 10^{-4}$$

## *I.* *W-annihilation and Penguin:*

$$B(B^+ \to \pi^+ K^{*0}) = 1.37\, B(B^+ \to K^+ \varphi) \qquad (95)$$
$$(1.01 \pm 0.01) \times 10^{-5} \qquad (1.21\,^{+0.09}_{-0.08}) \times 10^{-5}$$

$$B(B_c^+ \to D^+ K^{*0}) = 1.38\, B(B_c^+ \to D_s^+ \varphi) \qquad (96)$$
$$< 2 \times 10^{-7} \qquad < 4.4 \times 10^{-7}$$

## *J.* *W-internal, W-external & W- annihilation and Penguin:*

$$B(B^+ \to K^+ \rho^0) = 1.03\, B(B^+ \to K^+ \omega) \qquad (97)$$
$$(3.7 \pm 0.5) \times 10^{-6} \qquad (6.7 \pm 0.4) \times 10^{-6}$$

$$B(B^+ \to \eta \rho^+) = 1.22\, B(B^+ \to \eta' \rho^+) \qquad (98)$$
$$(7.0 \pm 2.9) \times 10^{-6} \qquad (11.8 \pm 2.7) \times 10^{-6}$$

## *K.* *W- internal emission, W- exchange & Penguin:*

$$B(B^0 \to \eta \rho^0) = 1.22\, B(B^0 \to \eta' \rho^0) \qquad (99)$$
$$< 1.5 \times 10^{-6} \qquad < 1.5 \times 10^{-6}$$

We note that almost all the relations obtained here are generally found to be consistent with experimental values. It may be remarked that the relations (57) to (88), involving only one weak process, remain unaffected by any change of phase of the decay amplitudes, which may arise due to elastic FSI. However, the relations (89) to (99), where two or more weak processes contribute, may be affected by the relative phase factor. Further, it may also be noted that the relations (59, 61, 64-65, 69, 88) follow from the QDS at isospin level, and hence are more reliable. However, branching fraction relations (70, 97) show deviation from the experimental values. Note that available branching fraction for $B_s^0$ decay are scaled with $B(\bar{b} \to B_s^0)$ [7].

## 5. Summary and Conclusions

Theoretically these decays are usually studied using the factorization scheme which expresses the decay amplitudes in terms of certain meson decay constants and meson-meson form factors. However, this scheme is unable to explain the experimental results, even after including hard QCD effects and possible phase differences. This may happen because of possible soft gluon exchange effects around the weak vertex, which enhance the contributions of W-exchange, W-annihilation and W-loop processes. Since these effects are not extractable from the first principles, we have investigated the *B*-meson decays employing the framework of QDS. Firstly, we have obtained decay amplitude relations among *B* to *PV* decays using *SU(2)*-isospin, *SU(2)*-U spin, and *SU(3)* for the QDS. Afterwards, relations among their corresponding branching fractions have been derived in Sec. 4, giving experimental results wherever available.

We make the following observations:

- The relations (64-65, 67-69, 71, 89, 91, 98) are consistent with the experimental data within the errors. So QDS seems to hold good for these hadronic weak decays of *B*-mesons. It may be noted that the relations (64-65, 69) follow from *SU(2)*- isospin QDS based.
- The relations (57-58, 66, 90, 95) show little deviation (~10%), which may be due to the scaling of experimental values of $B_s^0$ decays as given by Particle data group. Except for (66) rest of the relations are related through *SU(2)*- U spin, which being broken symmetry, may show slight deviation.

- Using the available branching fraction of the observed decays, we also predict the branching fraction of several decays in (59, 62, 72-76, 78, 81-84, 93-94). Explicitly $B(B_s^0 \to \eta_c \bar{K}^{*0}) = (0.05^{+0.03}_{-0.02}) \times 10^{-3}$, $B(B^+ \to \eta_c \rho^+) = (0.08^{+0.04}_{-0.03}) \times 10^{-3}$, $B(B_s^0 \to \bar{K}^0 \bar{D}^{*0}) = (7.09 \pm 2.36) \times 10^{-4}$, $B(B_s^0 \to D_s^- K^{*+}) = (2.85 \pm 0.47) \times 10^{-4}$, $B(B_s^0 \to K^+ D_s^{*-}) = (1.21 \pm 0.05) \times 10^{-4}$, $B(B_s^0 \to \eta_c \varphi) = (4.8 \pm 0.7) \times 10^{-4}$, $B(B^+ \to \pi^0 D_s^{*+}) = (1.1 \pm 0.2) \times 10^{-5}$, $B(B^0 \to \eta_c \rho^0) = (2.33 \pm 0.33) \times 10^{-5}$, $B(B_s^0 \to \pi^0 \bar{D}^{*0}) = (0.11 \pm 0.03) \times 10^{-5}$, $B(B_s^0 \to D^- \rho^+) = (0.3 \pm 0.1) \times 10^{-5}$, $B(B^0 \to \eta_c \omega) = (2.33 \pm 0.33) \times 10^{-5}$, $B(B_s^0 \to \pi^+ D^{*-}) = (1.31 \pm 0.18) \times 10^{-6}$, $B(B^+ \to D_s^+ \bar{K}^{*0}) = (0.12 \pm 0.06) \times 10^{-6}$, $B(B_s^0 \to K^- D^{*+}) = (1.3 \pm 0.2) \times 10^{-6}$, $B(B_s^0 \to \bar{D}^0 \omega) = B(B_s^0 \to \bar{D}^0 \rho^0) = (1.4 \pm 0.4) \times 10^{-6}$. Measurement of branching fractions of these decays would help to ascertain the validity of QDS.

- There are certain decay relations (60-61, 79-80, 96, 99) where only upper limits are available, for both sides; and few decay relations (63, 77, 85-88, 92) are obtained where upper limit are available for one of the decays. These results may be tested in future experiments.

- It may further be noted that the experimental branching fractions for the relations (70, 97) have the same order, though different magnitudes. This may happen due to the possible phase differences between the two or more weak processes involved in the decays. Also branching fraction relation (70) involve the decay of $B_s^0$ meson particularly, whose branching fractions are all scaled with $B(\bar{b} \to B_s^0)$, as mentioned in PDG [7]. The relation (97), where two or more weak processes contribute, may be affected by the relative phase factor and is a *SU(3)* based relation. This deviation may also point towards non-ideal mixing, or one may be tempted to hold *SU(3)* breaking responsible for this gap. However, looking at the large discrepancy we suggest new measurements for the decays involved in these relations to resolve the issue.

- Conventionally, W-exchange and W-annihilation diagrams are expected to be suppressed due to the helicity and color considerations. We find, in general, their contributions are not negligible, which indicates the significance of nonfactorizable soft gluon exchanges.

**References:**

1. M. Wirbel, “Description of weak decays of *D* and *B* mesons”, Prog. Part. Nucl. Phys **21**, 33 (1988).
2. A. J. Bevan *et al.,* “The Physics of the *B* Factories”, Eur. Phys. J. C **74**, 3026 (2014).

3. T. E. Browder, K. Honscheid and D. Pedrini, "Nonleptonic decays and lifetimes of *b* quark and *c* quark hadrons", Annu. Rev. Nucl. Part. Sci **46**, 395 (1996).
4. M. Wirbel, B. Stech and M. Bauer, "Exclusive Semileptonic Decays of Heavy Mesons", Z. Phys. C **29**, 637 (1985); M. Bauer, B. Stech and M. Wirbel, "Exclusive Nonleptonic Decays of $D$, $D_s$, and $B$ Mesons", *ibid.* C **34**, 103 (1987).
5. W. F. Wang and Z. J. Xiao, "The semileptonic decays $B/B_s \to (\pi, K)$ $(l^+ l^-, l\nu, l\nu\ \bar{\nu})$ in the perturbative QCD approach beyond the leading order", Phys. Rev. D **86**, 114025 (2012).
6. A. Ali, "*B* decays in the standard model: status and perspectives", Acta Phys. Polon. B **27**, 3529 (1996).
7. K. A. Olive *et al.* (Particle Data Group Collaboration), "Review of Particle Physics", Chin. Phys. C **38**, 090001 (2014) and 2015 update.
8. Maninder Kaur, Rohit Dhir, Avinash Sharma and R. C. Verma "Topological Diagram Analysis of Bottom Meson Decays Emitting Two Pseudoscalar Mesons", Phys. Part. Nucl. Lett **12**, 230 (2015).
9. J. P. Ma and C. Wang, "QCD factorization for quarkonium production in hadron collisions at low transverse momentum", Phys. Rev. D **93**, 014025 (2016).
10. K. Kolar, "Investigation of the factorization scheme dependence of finite order perturbative QCD calculations: searching for approximately ZERO factorization scheme", Eur. Phys. J. C **74**, 3123 (2014).
11. J. Sun, Y. Yang, Q. Chang and G. Lu, "Phenomenological study of the $B_c \to BP$, $BV$ decays with perturbative QCD approach", Phys. Rev. D **89**, 114019 (2014).
12. Q. Qin, Z. T. Zou, X. Yu, H. N. Li and C. D. Lu, "Perturbative QCD study of $B_s$ decays to a pseudoscalar meson and a tensor meson", Phys. Lett. B **732**, 36 (2014).
13. T. N. Pham, "Testing QCDF factorization with phase determinations in $B \to K\pi$, $K\rho$, and $K^*\pi$ decays", Phys. Rev. D **93**, 114019 (2016).
14. H. Y. Cheng, "Effects of final state interactions in hadronic *B*-decays", Int. J. Mod. Phys. A **21**, 650 (2006).
15. A. A. Petrov, "Implications of final state interactions in *B*-decays", J. Phys. Conf. Ser **69**, 012010 (2007).
16. C. D. Lu, Y. L. Shen and W. Wang, "Final State Interaction in $B \to KK$ Decays", Phys. Rev. D **73**, 034005 (2006).
17. R. C. Verma, "*SU(3)* flavor analysis of nonfactorizable contributions to $D \to PP$ decays", Phys. Lett. B **365**, 377 (1996).
18. F. M. A. Shamali and A. N. Kamal, "Nonfactorization in Cabibbo favored *B* decays", Phys. Rev. D **59**, 054020 (1999).
19. A. N. Kamal, A. B. Santra, T. Uppal and R. C. Verma, "Nonfactorization in hadronic two body Cabibbo favored decays of $D^0$ and $D^+$", Phys. Rev. D **53**, 2506 (1996).
20. K. K. Sharma, A. C. Katoch and R. C. Verma, "Nonfactorizable contributions to charm meson ($D \to PP$) decays", Z. Phys. C **76**, 311 (1997).
21. K. K. Sharma, A. C. Katoch and R. C. Verma, "Nonfactorizable contributions to weak $D \to PP$ decays", Z. Phys. C **75**, 253 (1997).
22. R. C. Verma, "Decay constants and form factors of *s*-wave and *p*-wave mesons in the covariant light-front quark model", J. Phys. G. **39**, 025005 (2012).
23. C. H. Chen and H. N. Li, "Nonfactorizable contributions to *B* meson decays into charmonia", Phys. Rev. D **71**, 114008 (2005).
24. R. Dhir and R. C. Verma, "Nonfactorizable contributions to hadronic decays of *D* mesons", J. Phys. G **34**, 637(2007).
25. M. Gronau, O. F. Hernandez, D. London and J. L. Rosner, "Broken *SU(3)* symmetry in two body *B* decays", Phys. Rev. D **52**, 6356 (1995).

26. C. W. Chiang, M. Gronau, J. L. Rosner and D. A. Suprun, “Charmless $B \rightarrow PP$ decays using flavor *SU(3)* symmetry”, Phys. Rev. D **70**, 034020 (2004).
27. C. W. Chiang, M. Gronau, Z. Luo, J. L. Rosner and D. A. Suprun, “Charmless $B \rightarrow VP$ decays using flavor *SU(3)* symmetry”, Phys. Rev. D **69**, 034001 (2004).
28. H. Y. Cheng, C. W. Chiang and A. L. Kuo, “A global analysis of two-body $D \rightarrow VP$ decays within the framework of flavor symmetry”, Phys. Rev. D **93**, 114010 (2016).
29. L. L. Chau, H. Y. Cheng and B. Tseng, “Analysis of two body decays of charmed baryons using the quark diagram scheme”, Phys. Rev. D **54**, 2132 (1996).
30. R. C. Verma and A. Sharma, “Quark diagram analysis of $b \rightarrow VV$ weak decays including smearing effects”, Phys. Rev. D **64**, 114018 (2001).
31. Rollin. J. Morrison and Michael. S. Witherell, “D Mesons”, Annu. Rev. Nucl. Part. Sct. **39**, 183 (1989).
32. K. Berkelman and S. L. Stone, “Decays of *B* mesons”, Annu. Rev. Nucl. Part. Sci **41**, 1 (1991).
33. T. E. Browder and K. Honscheid, “*B* mesons”, Prog. Part. Nucl. Phys **35**, 81 (1995).